\def\a{\alpha}\def\b{\beta}\def\d{\delta}
\def\g{\gamma}
\def\k{\kappa}\def\l{\lambda}\def\m{\mu}\def\n{\nu}\def\r{\rho}\def\s{\sigma}
\def\y{\eta}

\def\D{\Delta}

\def\de{\partial}
\def\id{\equiv}\def\mo{{-1}}\def\ha{{1\over 2}}
\def\({\left(}\def\){\right)}\def\[{\left[}\def\]{\right]}
\def\lra{\leftrightarrow}

\def\diag{{\rm diag}}

\def\mn{{\mu\nu}}\def\ij{{ij}}
\def\coo{coordinates }
\def\rep{representation }

\def\poi{Poincar\'e }
\def\cor{commutation relations }
\def\kp{$\k$-\poi }\def\km{$\k$-Minkowski }

\def\section#1{\bigskip\noindent{\bf#1}\smallskip}
\def\subsect#1{\bigskip\noindent{\it#1}\smallskip}
\def\nota{\footnote{$^\dagger$}}

\def\PL#1{Phys.\ Lett.\ {\bf#1}}\def\CMP#1{Commun.\ Math.\ Phys.\ {\bf#1}}

\def\PR#1{Phys.\ Rev.\ {\bf#1}}

\def\JMP#1{J.\ Math.\ Phys.\ {\bf#1}}

\def\JoP#1{J.\ Phys.\ {\bf#1}} \def\IJMP#1{Int.\ J. Mod.\ Phys.\ {\bf #1}}
\def\MPL#1{Mod.\ Phys.\ Lett.\ {\bf #1}}

\def\EJP#1{Eur.\ J.\ Phys.\ {\bf#1}}
\def\JHEP#1{JHEP\ {\bf#1}}\def\EPJ#1{Eur.\ Phys.\ J.\ {\bf#1}}

\def\ref#1{\medskip\everypar={\hangindent 2\parindent}#1}
\def\beginref{\begingroup
\bigskip
\centerline{\bf References}
\nobreak\noindent}
\def\endref{\par\endgroup}

\def\hx{\hat x}\def\hX{\hat X}\def\hM{\hat M}
\def\tM{\tilde M}\def\ab{{\a\b}}\def\rs{{\r\s}}
\def\cD{{\cal D}}\def\cF{{\cal F}}\def\cC{{\cal C}}\def\cO{{\cal O}}
\def\ot{\otimes}
\magnification=1200
{\nopagenumbers
\line{}
\vskip40pt
\centerline{\bf Unification of \km and extended Snyder spaces}
\vskip40pt
\centerline{{\bf S. Meljanac}\nota{e-mail:\ meljanac@irb.hr}}
\vskip5pt
\centerline {Rudjer Bo\v skovi\'c Institute, Theoretical Physics Division}
\centerline{Bljeni\v cka c. 54, 10002 Zagreb, Croatia}
\vskip10pt
\centerline{and}
\vskip5pt
\centerline{{\bf S. Mignemi}\nota{e-mail:\ smignemi@unica.it}}
\vskip5pt
\centerline {Dipartimento di Matematica, Universit\`a di Cagliari}
\centerline{via Ospedale 72, 09124 Cagliari, Italy}
\smallskip
\centerline{and INFN, Sezione di Cagliari}

\vskip40pt
{\noindent\centerline{\bf Abstract}}
\vskip5pt
In a recent paper, we have studied associative realizations of the noncommutative extended Snyder model, obtained by including the Lorentz generators (tensorial coordinates) and their conjugated momenta.
In this paper, we extend this result to also incorporate a covariant realization of the \kp spacetime.
We obtain the coproduct, the associative star product and the twist in a Weyl-ordered realization, to first order in
the noncommutativity parameters.
This could help the construction of a quantum field theory based on this geometry.

\vfil\eject}

\section{1. Introduction}

In the last decades, noncommutative geometry has become an important topic both in mathematics and in theoretical physics, where
it is considered a promising candidate for describing the structure of spacetime at the Planck scale. The mathematical framework
most suitable for physical applications is based on the formalism of Hopf algebra.
This structure is simultaneously an associative algebra and a coassociative coalgebra, satisfying some compatibility conditions [1],
which is apt to describe the quantization of spacetime and the deformation of the momentum composition laws.
Its physical applications are based on the observation that most theories of quantum gravity predict a granular structure of spacetime
at small scales, with the appearance of a minimal length, that renders the classical geometric description inadequate [2].

Among the most well-known models of this kind are the Snyder space [3,4] and the \kp model [5] with its associated \km spacetime [6].
Both of them are based on a deformation of the commutation relations of the position operators $\hx_i$: for Snyder space one
has\footnote{$^*$}{In this paper, Latin indices run from 0 to $N-1$, Greek indices from 0 to $N$.}
$$[\hx_i,\hx_j]=i\b\hM_\ij,\eqno(1)$$
where $\b$ is a constant with dimension length square, while the operators $\hM_\ij$
are the generators of Lorentz transformations. For \km geometry, in a covariant setting,
$$[\hx_i,\hx_j]=i(a_i\hx_j-a_j\hx_i),\eqno(2)$$
with $a_i$ a fixed vector with dimension of length. The length scale is usually identified with the Planck length.

While in \km the algebra of the position operators is closed, in the Snyder case a closed algebra is obtained only if one includes
the Lorentz generators with standard commutation relations as elements of the defining coordinate algebra. As a consequence, one cannot construct
a proper Hopf algebra using a representation of the $\hM_\ij$ in terms of the position and momentum operators, and in fact such attempt
gives rise to a noncoassociative bialgebra [4].

A way to obtain a coassociative Hopf algebra for the Snyder model was proposed in [7] and discussed in detail in [8], and is based
on the introduction of new degrees of freedom given by antisymmetric tensors, identified with the Lorentz generators $\hM_\ij$.
In this way, the Snyder algebra in $N$ dimensions becomes isomorphic to $so(1,N)$, and has been called extended Snyder algebra, to
distinguish it from the standard representation based solely on the $N$ position variables [4,7].

In this paper, using a result from [9],
we extend the formalism developed in ref.~[8] to include also the \km deformation and then achieve an unification of the
extended Snyder model with the \kp model. This will be obtained by deforming $so(1,N)$ to an algebra which leaves invariant
a suitable metric $g$ and will be denoted $so(1,N;g)$. We then consider a Weyl realization of this algebra in terms of an extended
Heisenberg algebra [8] and compute its coproduct, star product and twist.

We shall only consider the Weyl realization of the \kp Snyder algebra. Of course more general realizations can be obtained both
for Snyder [4,10] and \kp [11], that can easily be implemented in our formalism.
We should also remark that a unification of the Snyder and \kp models has been previously obtained in the standard formalism [12].
In that case, the resulting algebra is however not coassociative and the star products are not associative.

\section{2. \km Snyder spacetime and $so(1,N;g)$ algebra}
In ref.~[9] it was proposed a formalism for constructing $\k$-deformations of orthogonal groups by considering
transformations that leave invariant generic metric tensors. We apply this formalism to the extended Snyder model of
ref.~[8] to get an unification of it with \km spacetime.

Let us consider the Lorentz algebra $so(1,N)$ with generators $M_\mn$,
$$[M_\mn,M_{\r\s}]=i\l(\y_{\m\r}M_{\n\s}-\y_{\m\s}M_{\n\r}-\y_{\n\r}M_{\m\s}+\y_{\n\s}M_{\m\r}),\eqno(3)$$
where $\y=\diag(-1,1,\dots,1)$, and perform a change of basis, defining
$$\tilde M_\mn=(OMO^T)_\mn,\qquad g_\mn=(O\y O^T)_\mn,\eqno(4)$$
where $O$ is an $(N+1)\times(N+1)$ matrix with transpose $O^T$. Then,
$$\tM^T=-\tM,\qquad g^T=g,\qquad\det g\ne0,\eqno(5)$$
and
$$[\tM_\mn,\tM_{\r\s}]=i\l(g_{\m\r}\tM_{\n\s}-g_{\m\s}\tM_{\n\r}-g_{\n\r}\tM_{\m\s}+g_{\n\s}\tM_{\m\r}).\eqno(6)$$
We shall denote this algebra with general metric $g$ as $so(1,N;g)$.
In particular, we are interested in matrices $g_\mn$ of the form
$$g_\mn=\left(\matrix{-1&0&\dots&0&g_0\cr
0&1&\dots&0&g_1\cr&&\dots\cr0&0&\dots&1&g_{N-1}&\cr g_0&g_1&\dots&g_{N-1}&g_N\cr}\right),\eqno(7)$$
with det $g=-g_0^2+\sum_{i=1}^{N-1}g_i^2-g_N$.
If $g_N=0$, the algebra (6) reduces to the \kp algebra, if $g_0=\dots, g_{N-1}=0$, to the Snyder one.

In fact, we can introduce extended \coo $\hx_\mn=M_\mn$ and $\hX_\mn=\tM_\mn$, so that $\hX_\mn=(O\hx O^T)_\mn$.
The $\hx_\mn$ satisfy the standard $so(1,N)$ algebra, while we can split the relations (6) as
$$\eqalignno{&[\hX_{iN},\hX_{jN}]=i\l(g_{iN}\hX_{jN}-g_{jN}\hX_{iN}+g_{NN}\hX_{ij}),&\cr
&[\hX_{ij},\hX_{kN}]=i\l(g_{ik}\hX_{jN}-g_{jk}\hX_{iN}-g_{iN}\hX_{jk}+g_{jN}\hX_{ik}),&\cr
&[\hX_{ij},\hX_{kl}]=i\l(g_{ik}\hX_{jl}-g_{il}\hX_{jk}-g_{jk}\hX_{il}+g_{jl}\hX_{ik}).&(8)}$$
If we now define $\hX_{iN}=\k\,\hX_i$ and
$$g_\ij=\y_\ij,\quad g_{iN}=\k\, a_i,\quad g_{NN}=\k^2\b,\eqno(9)$$
we get the extended unified \km Snyder spacetime, with \cor
$$\eqalignno{&[\hX_i,\hX_j]=i\l(a_i\hX_j-a_j\hX_i+\b\hX_{ij}),&\cr
&[\hX_{ij},\hX_k]=i\l(\y_{ik}\hX_j-\y_{jk}\hX_i-a_i\hX_{jk}+a_j\hX_{ik}),&\cr
&[\hX_{ij},\hX_{kl}]=i\l(\y_{ik}\hX_{jl}-\y_{il}\hX_{jk}-\y_{jk}\hX_{il}+\y_{jl}\hX_{ik}).&(10)}$$
In this way we obtain an unification of the Snyder and \kp spacetimes using the formalism of extended \coo $\hX_\mn$.

\subsect{Example}
As an example, let us consider an $so(1,2;g)$ algebra. Generalization to higher dimensions is straightforward.
In this case,
$$g_\mn=\left(\matrix{-1&0&g_0\cr
0&1&g_1\cr g_0&g_1&g_2\cr}\right).\eqno(11)$$
The matrix $O$ is defined up to an $SO(1,2)$ transformation. In fact, given an $SO(1,2)$ matrix $R$, one has
$OR\y(OR)^T=O\y O^T=g$.
Assuming $g_2+g_0^2-g_1^2>0$ and defining $\r=\sqrt{g_2+g_0^2-g_1^2}$, one can write down a particularly simple
expression for $O$, namely
$$O_\mn=\left(\matrix{1&0&0\cr
0&1&0\cr-g_0&g_1&\r\cr}\right).\eqno(12)$$
Using (12), $\hX_\mn$ can then be written in terms of $\hx_\mn$ as
$$\hX_{01}=\hx_{01},\qquad\hX_{02}=\r\,\hx_{02}+g_1\,\hx_{01},\qquad\hX_{12}=\r\,\hx_{12}+g_0\,\hx_{01}.\eqno(13)$$


\section{3. Weyl representation of $so(1,N;g)$ with general metric}
Consider the generalized Heisenberg algebra
$$[x_\mn,x_{\r\s}]=[p^\mn,p^{\r\s}]=0,\qquad[x_\mn,p^{\r\s}]=i(\d_\m^{\ \r}\d_\n^{\ \s}-\d_\m^{\ \s}\d_\n^{\ \r}),\eqno(14)$$
and define $X_\mn=(O\,x\,O^T)_{\m\n}$ and $P^\mn=(O^\ddagger\,p\,O^\mo)^\mn$, where $O^\ddagger=(O^\mo)^T$.
These variables still satisfy the \cor (14), but their indices are raised and lowered by means of the metric $g_\mn$.

We want now to find a \rep of the $\hX_\mn$ defined in sect.~1 in terms of the Heisenberg algebra generated by $X_\mn$ and $P_\mn$.
The $\hX_\mn$ satisfy the $so(1,N;g)$ algebra (6), that we write as
$$[\hX_\mn,\hX_{\r\s}]=i\l C_{\mn,\r\s}^{\qquad\ab}\,\hX_{\a\b},\eqno(15)$$
where the structure constants are given by
$$C_{\mn,\r\s}^{\qquad\ab}=\ha\Big[-g_{\n\r}(\d_\m^{\ \a}\d_\s^{\ \b}-\d_\m^{\ \b}\d_\s^{\ \a})+g_{\m\s}(\d_\r^{\ \a}\d_\n^{\ \b}-\d_\r^{\ \b}\d_\n^{\ \a})
-(\m\lra\n)\Big].\eqno(16)$$
The symmetry properties $C_{\mn,\r\s}^{\qquad\ab}=-C_{\n\m,\r\s}^{\qquad\ab}=-C_{\mn,\s\r}^{\qquad\ab}=-C_{\mn,\r\s}^{\qquad\b\a}$ $=-C_{\r\s,\mn}^{\qquad\ab}$ hold.

In general, if the operators $\hX_\mn$ generate a Lie algebra with structure constants $C_{\mn,\r\s}^{\qquad\ab}$,
the universal realization of $\hX_\mn$ in terms of the Heisenberg algebra (14), corresponding to Weyl-symmetric ordering, is given by [8,13,14]
$$\hX_\mn=X_{\a\b}\[{\l\,{\cal C}\over 1-e^{-\l\cal C}}\]_{\mn}^{\quad\ab},\eqno(17)$$
where $\cC_{\mn}^{\quad\ab}=-\ha\,C_{\mn\rs}^{\qquad\ab}P^{\r\s}$.
This realization enjoys the property
$$e^{ik^\mn\hX_\mn}\triangleright1=e^{ik^\mn X_\mn},\qquad k^\mn\in{\bf R},\eqno(18)$$
where 
the action $\triangleright$ is defined as
$$X_\mn\triangleright f(X_\ab)=X_\mn f(X_\ab),\qquad P^\mn\triangleright f(X_\ab)=-i{\de f(X_\ab)\over\de X_\mn}=[P^\mn, f(X_\ab)].\eqno(19)$$

From (17) it follows that the Weyl realization of $\hX_\mn$ in terms of the generalized Heisenberg algebra generated by $X_\mn$ and $P^\mn$ reads
$$\hX_\mn=X_\mn+{\l\over2}X_\ab\,{\cal C}_{\mn}^{\quad\ab}+{\l^2\over12}X_\ab\({\cal C}^2\)_{\mn}^{\quad\ab}+{\cal O}(\l^4).\eqno(20)$$
where
$$\eqalignno{&{\cal C}_{\mn}^{\quad\ab}=\ha\Big(\d_\m^{\ \a}P_\n^{\ \b}+\d_\n^{\ \b}P_\m^{\ \a}-(\a\lra\b)\Big),&\cr
&\({\cal C}^2\)_\mn^{\quad\ab}=\ha\Big(2P_\m^{\ \a}P_\n^{\ \b}+\d_\n^{\ \b}P_{\m\r}P^{\r\a}+\d_\m^{\ \a}P_{\n\r}P^{\r\b}-(\a\lra\b)\Big),&(21)}$$
and the indices are lowered by means of the metric $g_\mn$.

Inserting $\cal C$ in (20), we find up to order $\l^2$,
$$\hat X_\mn=\ X_\mn+{\l\over 2}\(X_{\m\a}P_\n^{\ \a}-X_{\n\a}P_\m^{\ \a}\)
-{\l^2\over 12}(X_{\m\a}P_{\n\b}P^{\a\b}-X_{\n\a}P_{\m\b}P^{\a\b}-2X_{\a\b}P_\m^{\ \a}P_\n^{\ \b}),\eqno(22)$$
and
$$\eqalignno{&[\hX_\mn,P^{\r\s}]=i(\d_\m^{\ \r}\d_\n^{\ \s}-\d_\m^{\ \s}\d_\n^{\ \r})+{i\l\over2}(\d_\m^{\ \r}P_\n^{\ \s}-\d_\n^{\ \r}P_\m^{\ \s}+\d_\n^{\ \s}P_\m^{\ \r}-\d_\m^{\ \s}P_\n^{\ \r})\cr
&-{i\l^2\over 12}(\d_\m^{\ \r}P_{\n\a}P^{\s\a}-\d_\m^{\ \s}P_{\n\a}P^{\r\a}-\d_\n^{\ \r}P_{\m\a}P^{\s\a}+\d_\n^{\ \s}P_{\m\a}P^{\r\a}+2P_\m^{\ \r}P_\n^{\ \s}-2P_\n^{\ \r}P_\m^{\ \s}).\cr&&(23)}$$

Using the decomposition of $\hX_\mn$ introduced in sect.~1 and defining $X_i={1\over\k}X_{iN}$ and $P_i=\k P_{iN}$, we can then write
$$\eqalignno{\hX_i&=X_i+{\l\over 2}\Big(X_kP_i^{\ k}-\b X_{ik}P^k-a_iX_kP^k+a_kX_iP^k+a_jX_{ik}P^{jk}\Big)+\cO(\l^2),&\cr
\hat X_\ij&=X_\ij+{\l\over 2}\Big(X_iP_j+X_{ik}P_j^{\ k}+a_iX_{jk}P^k-(i\lra j)\Big)+\cO(\l^2).&(24)}$$

\vfill\eject

\section{4. Coproduct and star product in Weyl realization}
Formulae for coproduct and deformed addition of momenta can be deduced from those of ref.~[8], with the difference that now the sums are performed with the curved metric (7)
instead of the flat metric. In particular, defining
$$e^{{i\over2}k^{\mn}\hX_{\mn}} e^{{i\over2}q^\rs\hX_{\rs}}=e^{{i\over2}(k^{\mn}\oplus q^\mn)\hX_{\mn}}\id e^{{i\over2}\cD^\mn(k,q)\hX_\mn},\eqno(25)$$
one has
$$\cD^\mn(k^\ab,q^\ab)=\ k^\mn+q^\mn-{\l\over2}\Big(k^{\m\a}q^\n_{\ \a}-k^{\n\a}q^\m_{\ \a}\Big)+\cO(\l^2).\eqno(26)$$
The coproduct $\D P^\mn$ is then
$$\D P^\mn=\cD^\mn(P^\mn\otimes1,1\otimes P^\mn)=\ \D_0P^\mn-{\l\over2}\Big(P^{\m\a}\otimes P^\n_{\ \a}-P^{\n\a}\otimes P^\m_{\ \a}\Big)+\cO(\l^2),\quad\eqno(27)$$
where $\D_0P^\mn=P^\mn\otimes1+1\otimes P^\mn$.
The coproduct (27) is coassociative. Generally, Lie-deformed quantum Minkowski spaces admit both Hopf algebra and Hopf algebroid structures [15] and coproduct of momenta are coassociative [14].

The functions $\cD^\mn(q,k)$ satisfy the symmetry properties
$$\cD^\mn(q,k)\big|_\l=\cD^\mn(k,q)\big|_{-\l}.\eqno(28)$$
Moreover,
$$e^{{i\over2}k^\mn X_\mn}\star e^{{i\over2}q^\rs X_\rs}=
e^{{i\over2}\cD^\mn(k,q)X_\mn}.\eqno(29)$$
This star product is associative.
$\cD^\mn$ can be decomposed as
$$\eqalignno{\cD^i(k,q)=&\ k^i+q^i-{\l\over2}\Big(k_kq^{ik}-k^{ik}q_k+a_k(k^kq^i-k^iq^k)\Big)+\cO(\l^2)&\cr
\cD^\ij(k,q)=&\ k^\ij+q^\ij-{\l\over2}\Big(k^{ik}q^j_{\ k}+\b\,k^iq^j+a_k(k^{ik}q^j+k^iq^{jk})-(i\lra j)\Big)+\cO(\l^2),\qquad&(30)}$$
with symmetry properties
$$\cD^i(q,k)\big|_\l=\cD^i(k,q)\big|_{-\l},\qquad\cD^\ij(q,k)\big|_\l=\cD^\ij(k,q)\big|_{-\l}.\eqno(31)$$
\bigbreak

\section{5. The twist for the Weyl realization}
In this section, we construct the twist operator at first order in $\l$, using the results of [8].
The twist is defined as a bilinear operator such that $\D h=\cF\D_0h\cF^\mo$ for each $h\in so(1,N)$.

The twist in a Hopf algebroid sense can be computed by means of the formula [16]
$$\cF^\mo\id e^F=e^{-{i\over2}P^\mn\ot X_\mn}e^{{i\over2}P^\rs\ot\hX_\rs}.\eqno(32)$$
Using the CBH formula one gets
$$F={i\over2}\,P^{\m\n}\ot(\hX_\mn-X_\mn)+\dots.\eqno(33)$$
Substituting (20) in (33), one obtains
$$F={i\l\over2}P^{\a\g}\ot X_{\a\b}P_\g^{\ \b}+\cO(\l^2).\eqno(34)$$

It is easy to check that
$$\cF\D_0P^\mn\cF^\mo=\D P^\mn,\eqno(35)$$
with $\D P^\mn$ given in (27).
In particular,
$$\eqalignno{\D P^i=&\ \D_0P^i-{\l\over2}\Big(P^k\ot P^i_{\ k}-P^{ik}\ot P_k+a_k(P^k\ot P^i-P^i\ot P^k)\Big)+\cO(\l^2),\cr
\D P^\ij=&\ \D_0P^\ij-{\l\over2}\Big(P^{ik}\ot P^j_{\ k}+\b\,P^i\ot P^j+a_k(P^{ik}\ot P^j+P^i\ot P^{jk})\cr
&-(i\lra j)\Big)+\cO(\l^2),\qquad&(36)}$$
where
$$\D_0P^i=P^i\otimes1+1\otimes P^i,\qquad\D_0P^\ij=P^\ij\otimes1+1\otimes P^\ij.\eqno(37)$$

\section{6. Final remarks}
Using the formalism introduced in ref.~[9], we have been able to unify the extended Snyder model with the \km spacetime.
In contrast with the unified model proposed in ref.~[12], the resulting coalgebra is coassociative,
but this is obtained at the price of introducing new tensorial variables, whose physical interpretation is not evident,
although they should be considered as internal degrees of freedom.

Our result could be useful for the construction of dispersion relations [17] and an associative field theory on a general $\kappa$-deformed Snyder spacetime.

\beginref

\ref[1] S. Majid, {\it Foundations of quantum group theory}, Cambridge Un. Press 1995.

\ref[2] S. Doplicher, K. Fredenhagen and J. E. Roberts, \PL{B331}, 39 (1994); S. Doplicher, K. Fredenhagen and J. E. Roberts, \CMP{172}, 187 (1995).

\ref[3] H.S. Snyder, \PR{71}, 38 (1947).

\ref[4] M.V. Battisti and S. Meljanac, \PR{D82}, 024028 (2010);
M.V. Battisti and S. Meljanac, \PR{D79}, 067505 (2009).

\ref[5] J. Lukierski, H. Ruegg, A. Novicki and V.N. Tolstoi, \PL{B264}, 331 (1991); J. Lukierski and  H. Ruegg, \PL{B329}, 189 (1994).

\ref[6] S. Zakrewski, \JoP{A27}, 2075 (1994); S. Majid and  H. Ruegg, \PL{B334}, 348 (1994).

\ref[7] F. Girelli and E. Livine, \JHEP{1103}, 132 (2011).

\ref[8] S. Meljanac and S. Mignemi, \PR{D102}, 126011 (2020).

\ref[9] A. Borowiec and A. Pachol, \EPJ{C74}, 2812  (2014).

\ref[10] S. Meljanac, D. Meljanac, F. Mercati, D. Pikutic, \PL{B766}, 181 (2017);
S. Meljanac, D. Meljanac, S. Mignemi and R. \v Strajn, \PL{B768}, 321 (2017).

\ref[11] S. Meljanac and M. Stoji\'c, \EJP{C47}, 531 (2006); S. Meljanac, S.~Kre\v si\'c-Juri\'c and M. Stoji\'c, \EJP{C51}, 229 (2007);
 D. Kova\v cevi\'c and S. Meljanac, \JoP{A45}, 135208 (2012).

\ref[12] S. Meljanac, D. Meljanac, A. Samsarov and M. Stoji\'c, \MPL{A25}, 579 (2010); S. Meljanac, D. Meljanac, A. Samsarov and M. Stoji\'c, \PR{D83}, 065009 (2011).

\ref[13] S. Meljanac, T. Martini\'c-Bila\' c and S. Kre\v si\'c-Juri\'c, \JMP{61}, 051705 (2020).

\ref[14] N. Durov, S. Meljanac, A. Samsarov and Z. \v Skoda, J. Algebra {\bf 309}, 318 (2007).

\ref[15] J. Lukierski, D. Meljanac, S. Meljanac, D. Pikuti\'c, and M. Woronowicz, \PL{B777}, 1 (2018); J. Lukierski, S. Meljanac, and M. Woronowicz, \PL{B789}, 82 (2019).

\ref[16] S. Meljanac, D. Meljanac, A. Pachol, D. Pikuti\'c, \JoP{A50}, 265201 (2017);
S. Meljanac, D. Meljanac, S. Mignemi and R. \v Strajn, \IJMP{A32}, 1750172 (2017).

\ref[17] D. Meljanac, S. Meljanac, S. Mignemi, R. \v Strajn, \PR{D99}, 126012 (2019).

\endref
\end